\documentclass[preprint,amsmath,amssymb]{revtex4}
\usepackage{graphicx}
\usepackage{dcolumn}
\usepackage{bm}

\begin{document}

\title{Modularity map of the network of human cell differentiation}

\author{Viviane Galv\~{a}o$^{1,2}$, Jos\'{e} G. V.  Miranda$^2$, Roberto
  F. S. Andrade$^{2}$, Jos\'{e} S. Andrade Jr.$^{3,4}$, Lazaros
  K. Gallos$^5$, Hern\'{a}n A. Makse$^{3,5}$}

\affiliation{$^1$Departamento de Ci\^{e}ncias Biol\'{o}gicas, Universidade
Estadual de Feira de Santana, 44036-900, Feira de Santana, Bahia, 
Brazil\\
$^{2}$Instituto de F\'{i}sica, Universidade Federal da Bahia,
40210-210, Salvador, Bahia, Brazil.
\\$^{3}$Departamento de F\'{i}sica, Universidade Federal do Cear\'{a},
Campus do Pici, 60455-760, Fortaleza, Cear\'a, Brazil.
\\
$^4$ IfB, HIF E12, ETH Honggerberg, 8093 Zurich, Switzerland\\
$^{5}$Levich Institute and Physics Department, City College of New York,
New York, New York 10031, USA
}

\date{\today}

\begin{abstract}

  {\bf Cell differentiation in multicellular organisms is a complex
    process whose mechanism can be understood by a reductionist
    approach, in which the individual processes that control the
    generation of different cell types are identified. Alternatively,
    a large scale approach in search of different organizational
    features of the growth stages promises to reveal its modular
    global structure with the goal of discovering previously unknown
    relations between cell types. Here we sort and analyze a large set
    of scattered data to construct the network of human cell
    differentiation (NHCD) based on cell types (nodes) and
    differentiation steps (links) from the fertilized egg to a crying
    baby.  We discover a dynamical law of critical branching, which
    reveals a fractal regularity in the modular organization of
    the network, and allows us to observe the network at different
    scales.  The emerging picture clearly identifies clusters of cell
    types following a hierarchical organization, ranging from
    sub-modules to super-modules of specialized tissues and organs on
    varying scales.  This discovery will allow one to treat the
    development of a particular cell function in the context of the
    complex network of human development as a whole. Our results point
    to an integrated large-scale view of the network of cell types
    systematically revealing ties between previously unrelated domains
    in organ functions.}
\end{abstract}

\maketitle

\newpage

The cell differentiation process plays a crucial role in the prenatal
development of multicellular organisms. Recent advances in the
research on stem cell properties and embryonic development have
uncovered several steps in the differentiation process
\cite{Sell2004,Kirschstein2001,Freitas1999,Alberts2002,Sadler2004,Vickaryous,valentine,shapiro}.
Single and multiple sequences of cell differentiation have been
identified through in-vivo observations of a particular embryo during
early stages of development and through pathology studies of
miscarriages during late stages of the process. While the
identification of each cell differentiation step has been the subject
of intense research, an integrated view of this complex process is
still missing.  Such a global view promises to reveal features
associated with the large-scale modular organization of the cell types
\cite{Sadler2004,Vickaryous,valentine,ravasz,girvan,amaral,caldarelli,shm}
with the purpose of discovering new functional modules between cell
types using novel theoretical network analysis for community detection
\cite{girvan,amaral,caldarelli}.  In this letter, we take advantage of
the current knowledge on the sequence of cell differentiation
processes, which is spread over a vast specialized literature
\cite{Sell2004,Kirschstein2001,Freitas1999,Alberts2002,Sadler2004,Vickaryous,
  Paxinos2004,Temple2001,Bianco2001,Chen2003,Janeway2001,Anglani2004,Horster1999,Herrick2004,Otto2002,
  Jessen2005,Nakashima2003,Santagati2003,Savage2003,Forge2002,Panteleyev2001}
(see the Supplementary Information SI-Table \ref{table1} and
references therein), to reveal and characterize the topological and
dynamical features associated with the network of human cell
differentiation (NHCD).

\section{Results}

We construct the NHCD by systematically gathering the scattered
information on the evolution of each cell type present in the embryo
and fetus from a predecessor with a higher degree of differentiation
potential into a more specialized type.
The process of cell differentiation is then mapped onto a complex
network which consists of 873 nodes connected through 977 edges. The
nodes in the network represent distinct cell types reported in the
literature
\cite{Sell2004,Kirschstein2001,Freitas1999,Alberts2002,Sadler2004,Vickaryous,
  Paxinos2004,Temple2001,Bianco2001,Chen2003,Janeway2001,Anglani2004,Horster1999,Herrick2004,Otto2002,
  Jessen2005,Nakashima2003,Santagati2003,Savage2003,Forge2002,Panteleyev2001}
and the edges represent the association between two cell types through
a differentiation event.

The initial steps of the NHCD are shown in the inset of
Fig. \ref{fig1}, while the resulting network structure is shown in the
main panel of Fig. \ref{fig1} (see also SI-Figs. \ref{fig3}a and
\ref{fig3}b).  The fertilized egg is followed by the ball stage, and
the formation of the primary germ cell layers.  Currently, it is known
that until the ball stage, cell division is symmetric and produces
further totipotent stem cells \cite{Sell2004}. These cells then give
rise to all the differentiated tissues of the organism as well as the
extra-embryonic tissues (placenta, umbilical cord, etc.). Moreover, in
the course of the entire process of organism formation, there is a
monotonic decrease in the differentiation potential (totipotent
$\rightarrow$ pluripotent $\rightarrow$ multipotent $\rightarrow$
unipotent cells) accompanied with an increase in cell
specialization.

Certain types of cells can be generated following more than one path
from the fertilized egg. This process generates some closed loops of
edges in the network.  The NHCD comprises 529 branches of different
lengths with each branch ending when the cell types do not undergo
further differentiation.  Note, however, that the most recent
compilation of cell types in normal, healthy, human adults done in
\cite{Vickaryous} reports only 407 cell types.  Therefore, not all
branch endpoints correspond to cell types in born humans.
Thus, not all 873 cell types are present in a human being. Among those
absent are the placenta cells that are generated from the fertilized
egg during embryo development, as well as other somatic cell types
that are important to control embryo and fetus development.  The cell
types that survive in a human are denoted by filled circles, while
non-surviving ones are indicated by empty circles.  The complete
collected data is listed in the Supplementary Information, including
an enumeration of cells and links between the cell types, their time
of appearance in days after fecundation ($T_a$), and the reference to
the publications reporting each link. To the best of our knowledge,
the structure identified here provides the most complete schematic
diagram of the human differentiation process to date.

It is visually apparent from Fig. \ref{fig1} that the NHCD has a
prominent modular structure. The continuous differentiation of cells
into more specialized functions naturally leads to the formation of
dense isolated clusters in the NHCD.  As a first approach to
understand this modular structure we cluster cell types in the network
of Fig. \ref{fig1} according to their known functions; different
colors indicate 19 functional modules extracted from the literature
$(C1-C19)$ (See SI-Table \ref{table1} and references therein.  The
largest communities were extracted from Refs.
\cite{Sell2004,Kirschstein2001,Freitas1999,Alberts2002,Sadler2004,Vickaryous,
  Paxinos2004,Temple2001,Bianco2001,Chen2003,Janeway2001,Anglani2004,
  Horster1999,Herrick2004,Otto2002,Jessen2005,Nakashima2003,Santagati2003,
  Savage2003,Forge2002,Panteleyev2001}).  There is, however, a certain
degree of arbitrariness in this modular structure as the separation of
the nodes into communities in our dataset is not unique. For instance,
community $C12$, the neural lineage, could be divided into two
sub-communities, representing the neural and the supporting (glial)
cells
\cite{Freitas1999,Kirschstein2001,Paxinos2004,Sadler2004,Sell2004,Temple2001,Vickaryous}. On
the other hand, the neural system module could be merged with the eye
system module \cite{Paxinos2004,Sadler2004,Sell2004,Vickaryous} on a
larger scale, since they have a common ancestral cell type.
Therefore, a finer or coarser community structure can be extracted
from the data when we look at the whole network at different scales of
observation; a novel module-detection algorithm is needed to identify
these communities in a systematic way.

Graph theoretical concepts allow us to unravel the scale dependence of
the modular structure of the NHCD. Graph theory \cite{caldarelli}
defines the distance between two nodes (also called the chemical
distance) as the number of links along the shortest path between the
nodes in the network. We use this notion to propose a community
detection algorithm that identifies modules of size $\ell$ composed of
highly connected cell types.  The algorithm finds the optimal tiling
of the network with the smallest possible number of modules, $N_B$, of
size $\ell$ \cite{shm} (each node is assigned to a module or box and
all nodes in a module are at distance smaller than $\ell$). This
process results in an optimization problem which can be solved using
the box-covering algorithm explained in Fig.\ref{modules}a, Materials
and Methods Section \ref{algorithm} and reported in \cite{jstat} as
the Maximum Excluded Mass Burning algorithm (MEMB, the algorithm can
be downloaded from
\url{http://lev.ccny.cuny.edu/~hmakse/soft_data.html}).  The
requirement of minimal number of modules to cover the network $(N_B)$
guarantees that the partition of the network is such that each module
contains the largest possible number of nodes and links inside the
module with the constraint that the modules cannot exceed size
$\ell$. This optimized tiling process gives rise to modules with the
fewest number of links connecting to other modules implying that
the degree of modularity, defined by
\cite{girvan,amaral,caldarelli,lazaros}
\begin{equation}
{\cal M}(\ell) \equiv \frac{1}{N_B} \sum_{i=1}^{N_B} \frac{L_i^{\rm
    in}}{L_i^{\rm out}} ,
\label{mo}
\end{equation}
is maximized. Here $L_i^{\rm in}$ and $L_i^{\rm out}$ represent the
number of links that start in a given module $i$ and end either within
or outside $i$, respectively.
Large values of $\cal M$ $(L_i^{\rm out}\to 0)$ correspond to a higher
degree of modularity.  The value of the modularity of the network
$\cal M$ varies with $\ell$, so that we can detect the
dependence of modularity on different length scales, or equivalently
how the modules themselves are organized into larger modules that
enhance the degree of modularity.


For a given $\ell$, we obtain the optimal coverage of the network with
$N_B$ modules (we use the MEMB algorithm \cite{jstat} explained in
Fig. \ref{modules}a and Materials and Methods). Analysis of the
modularity Eq. (\ref{mo}) in Fig. \ref{fig7}a reveals a monotonic
increase of ${\cal M}(\ell)$ with a lack of a characteristic value of
$\ell$.  Indeed, the data can be approximately fitted with a power-law
functional form:
\begin{equation}
{\cal M}(\ell) \sim \ell^{d_M} ,
\label{modular}
\end{equation}
which is detected through the modularity exponent $d_M$.  We
characterize the network using different snapshots in time and we find
that $d_M \simeq 2.0$ is approximately constant over the time
evolution (Fig. \ref{fig7}a).  This value reveals a considerable
degree of modularity in the entire system (for comparison, a random
network has $d_M=0$ and a uniform lattice has $d_M=1$ \cite{lazaros}),
as evidenced by the network structure in Fig.~ \ref{fig1}. The lack of
a characteristic length-scale in the modularity shown in
Fig. \ref{fig7}a
suggests that the modules appear at all length-scales, i.e. modules
are organized within larger modules in a self-similar way, so that the
inter-connections between those clusters repeat the basic modular
character of the entire NHCD.  Thus, the NHCD remains statistically
invariant when observed at different scales.
Varying the module size $\ell$ yields the scaling relation for the
number of modules (Fig.~\ref{fig7}b): 
\begin{equation}
N_B(\ell) \sim \ell^{-d_B},
\label{db}
\end{equation}
where $d_B$ represents the fractal dimension of the network
\cite{shm}.  We find that the fractal character of the modules is
established at the early stages, yielding $d_B\simeq 1.4$ as early as
30 days (Fig.~\ref{fig7}b).  As the network evolves, the fractal
dimension increases slightly and finally reaches $d_B\simeq 1.9$.

The significance of Eq. (\ref{modular}) is that the modules need to be
interpreted at a given length-scale.  Figure \ref{modules}b shows an
example of such hierarchical organization [Fig. \ref{modules}c and
SI-Fig. \ref{full} show the full modular structure, while a list of
detected modules appears in the SI Appendix]. Three types of communities of
cell types are clearly identified in Fig. \ref{modules}b as we change
$\ell$. {\it (i) The known functional modules:} The entire eye lineage
\cite{Paxinos2004,Sadler2004,Sell2004,Vickaryous} is detected as a
single module by the box-covering algorithm at $\ell=11$, while the
entire neural lineage
\cite{Freitas1999,Kirschstein2001,Paxinos2004,Sadler2004,Sell2004,Temple2001,Vickaryous}
appears at $\ell=15$. Finer and coarser novel modules are identified
by the algorithm.  {\it (ii) Sub-modules:} At $\ell=11$ the neural
lineage is split into the main neural and the supporting glial cell
modules, while for $\ell =7$ sub-modules are identified in the eye
system.  {\it (iii) Super-modules:} When we increase the length to
$\ell=19$, the eye and neural system form a single super-module.
Thus, each cell type is connected to other types according to which
groups of nodes of all sizes self-organize following a single
principle. This property allows us to renormalize the network
\cite{shm} by replacing each detected module by a single supernode to
identify the network of modules as shown in Fig.~\ref{modules}c.
Following the evolution and inter-dependence of these super-modules,
as seen in Fig.~\ref{modules}c, identifies families of cell types at
varying scales. This modularity map is useful in proposing future
research ties between previously unrelated domains in organ functions.

The dynamics leading to such a structure can be unraveled by the study
of the NHCD as a growth process.  The knowledge of the time of
appearance of each cell type, $T_a$, makes it possible to follow the
cumulative growth of the embryo and fetus in terms of the total number
of cell types at time $t$, $N(t)$ as well as the number of cell types
that eventually survive in the organism (Fig. \ref{fig4}a).  As
expected, surviving cells emerge in the later stages of the gestation
period. However, the difference between the total and the surviving
number of cell types indicates that generation of new types of
non-surviving cells takes place even during the final gestation
months.


The increase of the network size, $N(t)$, is initially approximately
exponential and after $t_*=40$ days
changes into a slower growth (Fig. \ref{fig4}a).  Only a small
percentage of the nodes grow within a given time interval, so that the
network activity is focused in a small number of them at a given time.
The number of nodes that differentiate at a given time are shown in
Fig.~\ref{fig4}b. We observe an activity that increases monotonically
up to around $t_*=40$ days and then drops to lower values.  The
cross-over time $t_*=40$ days observed in Figs. \ref{fig4}a and
\ref{fig4}b separates two regimes of growth and approximately
corresponds to the time below which most of the cells have a plastic
characteristic (i.e., the capability to differentiate) and above which
they start to become functional \cite{Sell2004}. Interestingly, the
two regimes observed in $N(t)$ merge into a single universal
functional curve when we replot $N(\ell_{N1})$ as a function of the
chemical distance to the fertilized egg, $\ell_{N1}$
(Fig. \ref{fig4}c). This result suggests that the topological distance
in the network $\ell_{N1}$ is the natural variable to characterize the
growth process in a universal form rather than the time. The dynamic
of $N(\ell)$ follows a typical logistic (Verhulst) process of
population growth where the rate of growth is restricted by
environmental limitations: 
\begin{equation}
\frac{dN}{d\ell}=r N [1-\frac{N}{N_f}],
\end{equation}
 with solution,
\begin{equation}
  N(\ell)=N_f \frac{\exp(r \ell)}{N_f+(\exp(r\ell)-1)}
\end{equation}
 (see the fitting in
Fig. \ref{fig4}c) where $N_f$ is the final number of cell types and
$r=0.65$ is the growth rate of cell types.

Analysis of the network connectivity reveals that the average number
of links per node in the final stages of the entire NHCD is $\langle k
\rangle = 2.24$ (Fig.~\ref{fig4}d).
Even though $\langle k\rangle \approx 2$, there is a broad degree
distribution (scale-free \cite{caldarelli}, $P(k)\sim k^{-\gamma}$,
$\gamma\simeq 3.0$, SI-Fig.~\ref{fig5}).  This implies that there is
always a small number of crucial cell types that differentiate much
more than the others, a fact that agrees with evidence on the
existence of a few cells with large plasticity potential. As this
potential is rapidly lost after 40 days, cell types change their
development ability in favor of the organism life maintenance.

The fact that the average degree is close to 2 implies that the
dynamical evolution of NHCD can be described by a critical branching
process where every node has a certain probability of generating
offsprings,
in which case the critical condition for the branching to continue is
$\langle k\rangle = 2$ \cite{athreya}. This effectively means that
each node needs to give at least one descendant in order for the
network to keep growing. If $\langle k\rangle < 2$, the growth would
stop early, while for $\langle k\rangle > 2$ the growth would be
faster than exponential.

The network reaches the condition of criticality, $\langle k\rangle
\approx 2$, at around $t_*=40$ days (Fig.~ \ref{fig4}d) in conjunction
with the transition from plasticity to functional behavior. After
this, the average degree remains just above criticality to sustain a
growth rate that guarantees the network survival.  The majority of the
nodes propagate the growth in a single line, but there are nodes which
generate significantly more descendants to generate the diversity
implied by the power-law distributions of degree and modularity. 

\section{Discussion}

In summary, we present the first large-scale study of the prenatal
evolution of the human cell differentiation process from the
fertilized egg to a developed human.
The process of human cell differentiation can be mapped onto a complex
network composed of cell types and differentiation steps.
This mapping allows us to study the cell differentiation process with
state of the art network theory for community detection with the goal
of identifying hitherto unknown functional relations between cell
types.

We discover a dynamical law of critical branching explaining the
emergence of the network topology, which reveals a novel
scale-invariant modular structure of the network of cell types.  
The self-similar modular features evidenced in Figs. \ref{fig1},
\ref{modules} and \ref{fig7} are established early in the process and
remain invariant during the evolution of the NHCD, although the
network size changes significantly.

Using this law, we are able to observe the network at different
scales. The emerging picture clearly identifies clusters of cell
types, or modules, and their connectivity to other modules within its
own and other functions.  The resulting hierarchical organization
consists of sub-modules, known biological functions and super-modules
of specialized tissues and organs emerging on varying scales. This
discovery is useful in proposing future research ties between
previously unrelated domains in organ functions in a systematic
way. Furthermore, this information could be of importance in providing
predictions of functional attributes to a number of identified modules
of cell types in the NHCD.


\section{\bf Materials and Methods}

\label{algorithm}

{\bf Module detection algorithm}

The detection of modules or boxes in our work follows from the
application of the box-covering algorithm \cite{shm,jstat} at
different length scales.  The algorithm can be downloaded at
\url{http://lev.ccny.cuny.edu/~hmakse/soft_data.html}.  In box
covering we assign every node to a module, by finding the minimum
possible number of boxes, $N_B(\ell)$, that cover the network and
whose diameter (defined as the maximum distance between any two nodes
in this box) is smaller than $\ell$. These boxes are characterized by
the proximity between all their nodes, at a given length scale.
Different values of the box diameter $\ell$ yield boxes of different
size.  These boxes are identified as modules which at a smaller scale
$\ell$ may be separated, but merge into larger entities as we increase
$\ell$.

In this work we implement the Maximum Excluded Mass Burning (MEMB)
algorithm from \cite{jstat} for box covering.  The algorithm uses the
basic idea of box optimization, where we require that each box should
cover the maximum possible number of nodes, and works as follows: For
a given $\ell$, we first locate the optimal `central' nodes which will
act as the origins for the boxes.  This is done by first calculating
the number of nodes (called the mass) within a diameter $\ell$ from
each node.  The node that yields the largest mass is marked as a
center.  Then we mark all the nodes in the box of this center node as
`tagged'.  We repeat the process of calculating the mass of the boxes
starting from all non-center nodes, and we identify a second center
according to the largest remaining mass, while nodes in the
corresponding box are `tagged', and so on.  When all nodes are either
centers or `tagged' we have identified the minimum number of centers
that can cover the network at the given $\ell$ value. Starting from
these centers as box origins, we then simultaneously burn the boxes
from each origin until the entire network is covered, i.e. each node
is assigned to one box (we call this process burning since it is
similar to burning algorithms developed to investigate clustering
statistics in percolation theory \cite{caldarelli}).  In
Fig.~\ref{modules}a of the main text we show how box-covering works
for a simple network at two different $\ell$ values.

This algorithm is driven by the proximity between nodes and the
maximization of the mass associated with each module center
\cite{shm,jstat}. Thus it detects boxes that maximize modularity,
Eq.~(\ref{mo}). In the case of MEMB we have the additional benefit of
detecting modules at different scales, so that we can study the
hierarchical character of modularity, i.e. modules of modules, and we
can detect whether modularity is a feature of the network that remains
scale-invariant.

The fractal dimension $d_B$ of a complex network is an exponent
that determines how the mass (equivalently: the number of nodes)
around any given node scales with the length, which in networks
corresponds to the shortest distance between two nodes.
In order to numerically measure this exponent we optimally cover the
network with boxes using the MEMB algorithm. A box is a set of nodes
where all distances $\ell_{ij}$ between any two nodes $i$ and $j$ in
this set are smaller than a given value of $\ell$, the box size.
Although there is a large number of coverings, for every value of
$\ell$ we want to find the one which gives the smallest possible
number of boxes, $N_B(\ell)$. Varying $\ell$ then yields the scaling
relation Eq. (\ref{db}). A finite fractal dimension reveals
fundamental organizational principles of the underlying network,
namely a self-similar structural character, where the network is built
in a similar way even though we observe it at different
length-scales. The boxes that are identified through this process
correspond to the modules at varying scales.

%
%
%

\clearpage

\bibliographystyle{prsty}

\noindent{\bf Acknowledgments:} We thank B. Bruji\'c, B. Dubin-Thaler,
H.D. Rozenfeld, S. Havlin, T. Rattei, M. Sigman and A. M. Andrade and
J. M. M. Andrade for valuable discussions.  This work was supported by
National Science Foundation Grants SES-0624116 and EF-0827508. We
thank the Brazilian agencies CNPq, CAPES, FAPESB and FUNCAP, and the
EU FP7 Neuronano project (NMP4-SL-2008-214547) for financial support

\clearpage

\begin{figure}
\begin{center}
%
\includegraphics*[width=16cm]{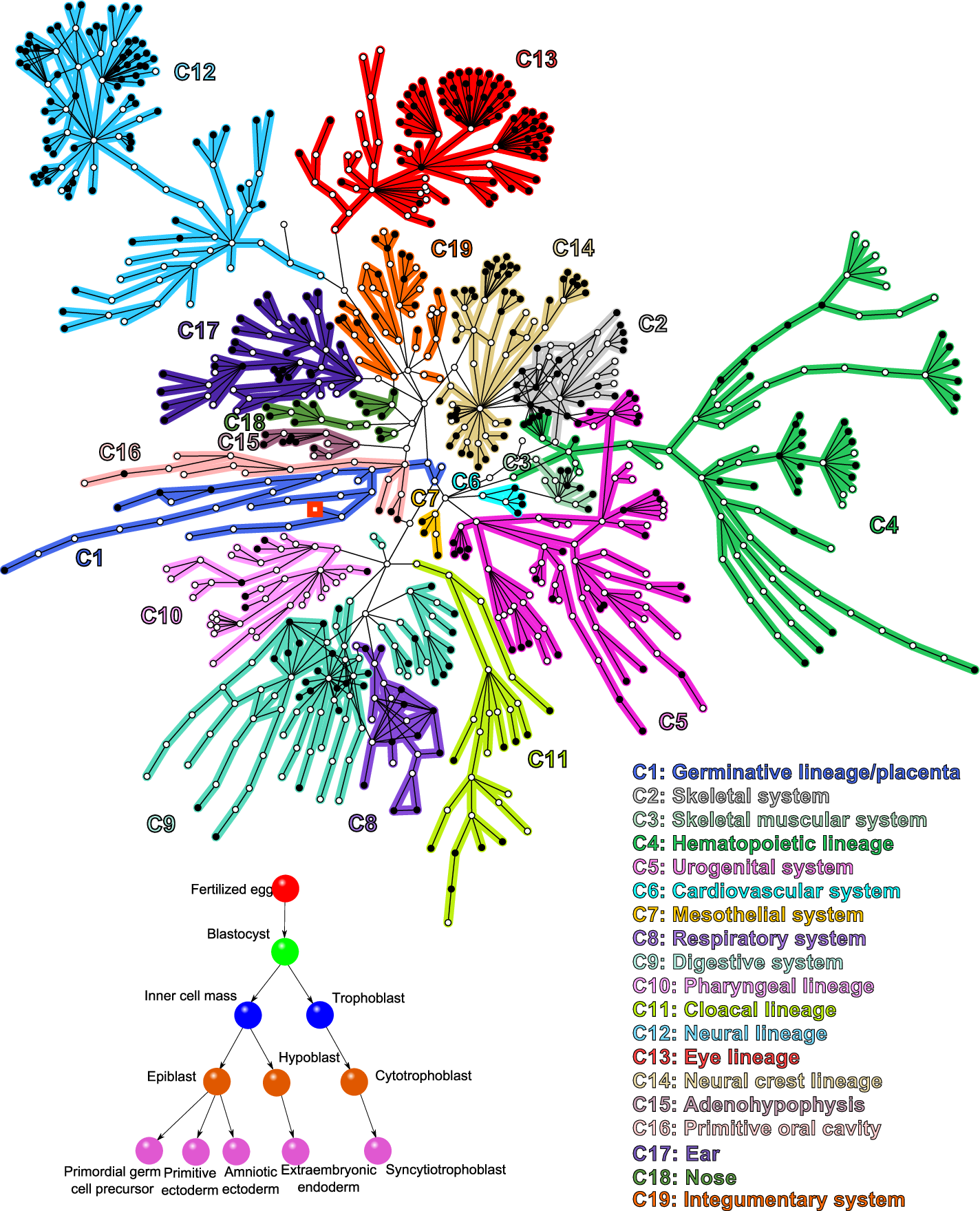}
\end{center}
\caption{}  \label{fig1}
\end{figure}

{\bf Fig. \ref{fig1}.} {\bf Complex network representation of the human
cell differentiation process.} The first steps of the NHCD construction
are shown in the inset of this figure.  These steps, known to also be
present in the formation of the majority of multicellular organisms,
include the first cleavage of a fertilized egg, which is subsequently
followed by the ball stage and the formation of primary germ cell
layers, namely, the ectoderm, mesoderm, and endoderm. The fertilized
egg is a totipotent stem cell. The blastocyst, in turn, gives rise to
both trophoblast and inner cell mass. These two cells further
differentiate into other types of cells, and so on. Following the
above process until the fetus is fully developed yields the complex
network shown in this figure. Each node, plotted as a circle,
corresponds to a cell type and the edges to a differentiation
step. The entire network originates from the fertilized egg (denoted
by a red square) and leads to the specialized cells of a developed
human.  Filled circles correspond to nodes that survive at the end of
the development process, while empty circles correspond to
non-surviving cell types.  Nodes in communities of known functions
from the literature are indicated by different colors, except for
those cell types with no functional annotation (see SI-Table
\ref{table1} for association to the known functions $C1$ to $C19$
extracted from the literature).

{\bf Fig. \ref{modules}.} {\bf Detection of modules and the network of
modules at different scales.}  {\bf a,} Demonstration of the
box-covering algorithm for a schematic network, following the Maximum
Excluded Mass Burning algorithm in \cite{shm,jstat} (see SI-Section
\ref{algorithm} for full details).  We cover the network with the
smallest possible number of boxes for a given $\ell$ value. This is
done in a two-stage process: {\it (i)} We detect the smallest possible
number of box origins (shown with cyan color) that provide the maximum
number of nodes (mass) in each box, according to the following
optimization algorithm: We calculate the mass associated with each
node, and pick the first center as the node with largest mass and mark
the nodes in this box as `tagged'. We repeat the process from the
remaining non-center nodes to identify a second center with the
highest mass, and so on. {\it (ii)} We build the boxes through
simultaneous burning from these center nodes, until the entire network
is covered with boxes.  For example, at $\ell=3$ there are four boxes,
where the maximum distance between any two nodes in a box is smaller
than $\ell$. Similarly, we can cover the same network with two boxes
at $\ell=6$. These two boxes are the result of merging two of the four
boxes at $\ell=3$.  {\bf b,} Detail of NHCD modules detected by the
above box-covering algorithm for two particular functions.  The
algorithm detects a hierarchy of sub-modules, known functions and
super-modules of size $\ell$ plotted in different colors. We show the
identified modules corresponding to $C12$-neural system and $C13$-eye
system (full structure is in Fig. \ref{modules}c and
SI-Fig. \ref{full}), which first appear at $\ell=15$ and $\ell=11$,
respectively.  At other scales the box-covering algorithm detects new
functional relations between cell types expressed in the obtained sub
and super-modules. For instance, at $\ell=11$ the neural lineage is
further divided into two sub-modules, while at $\ell=19$ the two
functions merge into a super-module.  {\bf c,} The network of modules
at different $\ell$ values, as detected through the box-covering
algorithm.  Every node corresponds to one of the three following
types, in terms of increasing scale: {\it (i)} Sub-modules (small grey
dots), which are fractions of a fully functional module, {\it (ii)}
Known functional biological modules (colored circles), whose color
corresponds to the functions $C1$-$C19$, and {\it (iii)} Super-modules
(pie-charts), which represent the union of more than one known
functional module, described by the colors of the pie-chart.  
The links that stem from known functional modules and super-modules
are shown in red, and they progressively span the entire network as we
increase $\ell$.

{\bf Fig. \ref{fig7}.} {\bf Modular properties of the NHCD.}  {\bf a,}
Degree of modularity of the network, ${\cal M}(\ell)$ at different
times $T_a$ (indicated in the figure) as a function of the scale of
observation, $\ell$.  {\bf b,} Number of boxes/modules, $N_B$, versus
the size of the modules $\ell$ identified by the box-covering
algorithm for different networks at time $T_a$.

{\bf Fig. \ref{fig4}.} {\bf Growth properties of the NHCD.} {\bf a,} Number
of cell types in the network, $N(t)$, as a function of time.  We find
precise information about the appearance time $T_{a}$ for 782 among
the 873 cell types. Those cells with missing appearance time have not
been taken into account in this plot. Also shown are the time
evolution of the number of surviving and non-surviving cells.  
{\bf b,} Number of nodes whose degree increases at time $t$
(red histogram) and number of new links appearing in the network (blue
histogram) as a function of time. If all nodes were giving just one
child then the two histograms would coincide.  Inset: The average
number of new links per node at a given time can be found by dividing
the two histograms in the main plot.  This plot shows how intense is
the activity at that particular time. Despite the variation in
activity, the new connections average around 1, which gives a critical
branching ratio of $\langle k \rangle \simeq 2$.  {\bf c,} Number of
cell types versus the chemical distance to the first node,
$\ell_{N1}$. This distance is only determined by the connections
between the cell types, and is not influenced by the appearance time,
so that we include all 873 cell types. {\bf d,} Average degree
$\langle k \rangle$ of the network as a function of time showing that
the network achieves the condition of critical branching process
$\langle k \rangle\approx 2$ at around $t_*=40$ days.

\clearpage

\clearpage

\begin{figure}
\begin{center}
\includegraphics*[height=11cm,angle=0]{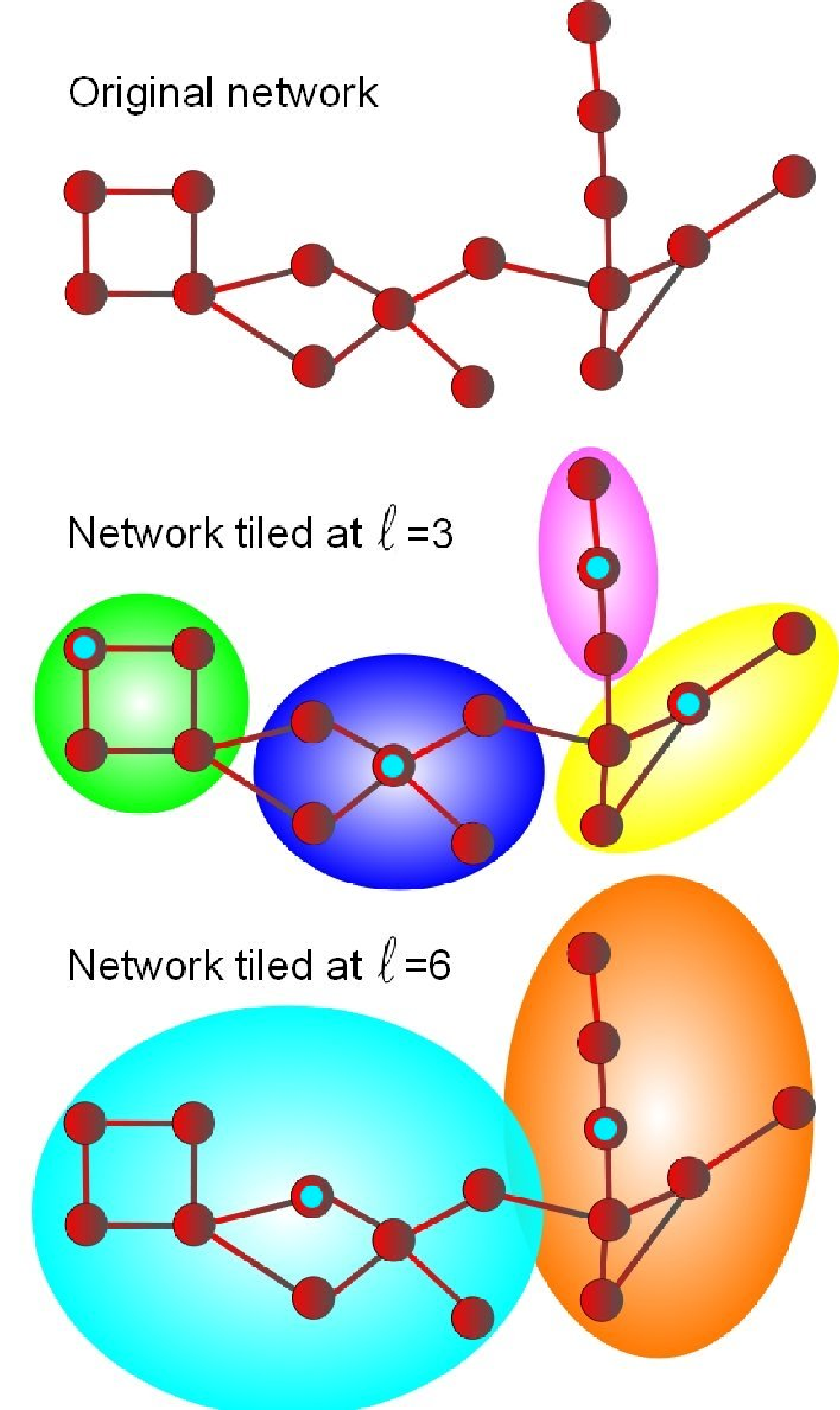}

{\bf Fig. 2a}

\vspace{1cm}

\includegraphics[width=15.5cm]{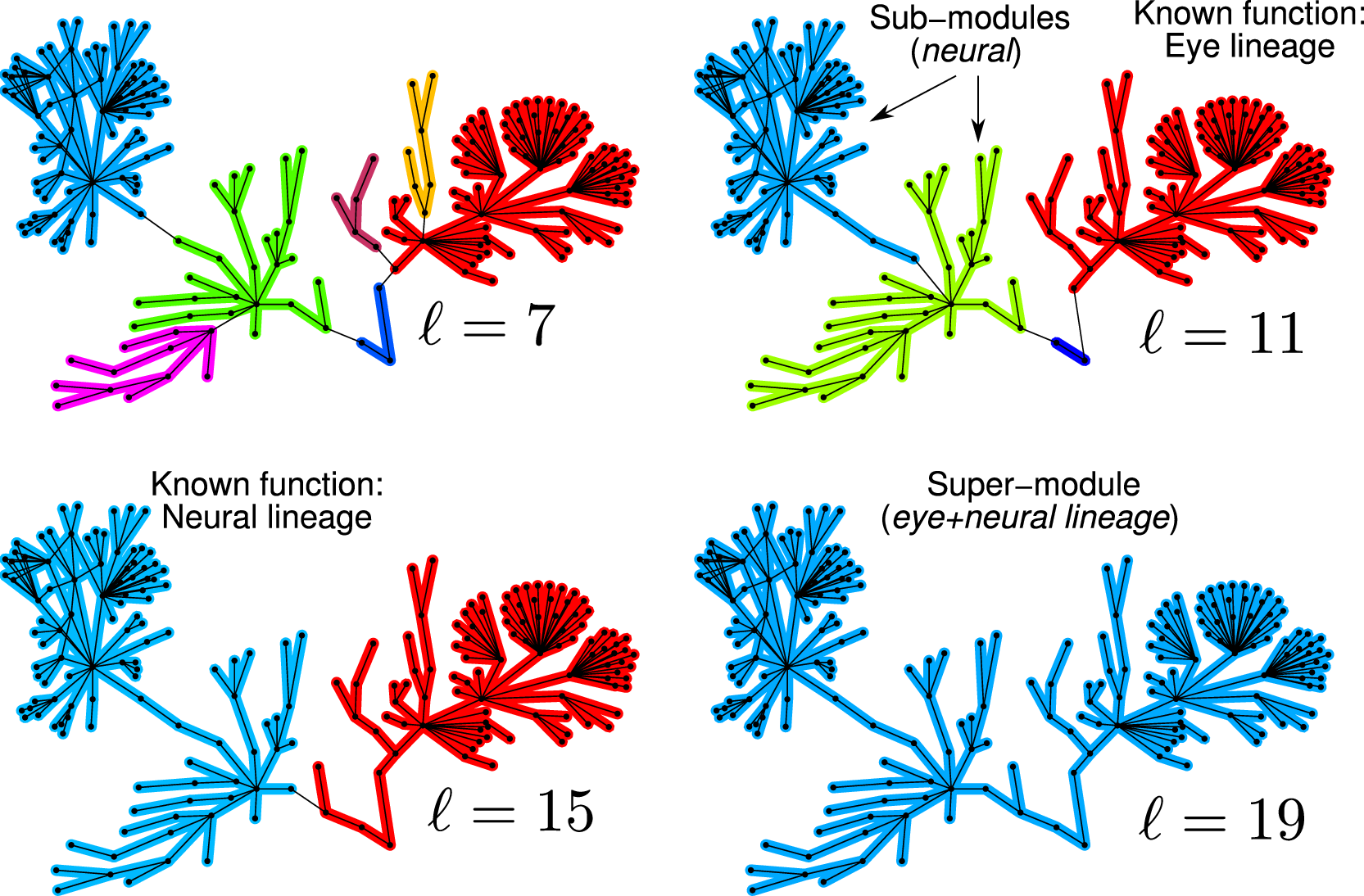}

{\bf Fig. 2b}

\end{center}
\end{figure}
\clearpage
\begin{figure}
\begin{center}
\includegraphics[width=15.5cm]{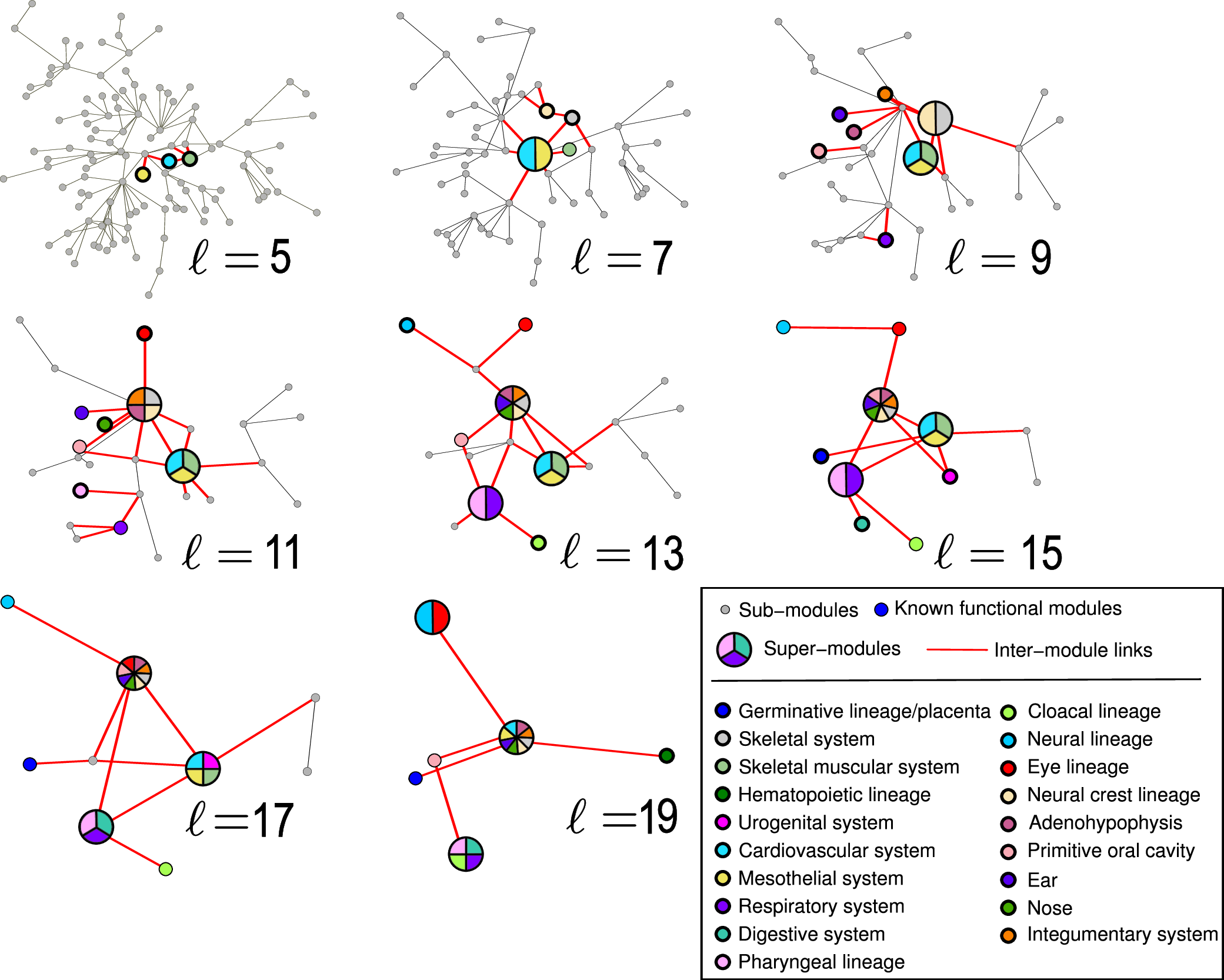}

{\bf Fig. 2c} 
\end{center}
\caption{}  
\label{modules}
\end{figure}

\clearpage

\begin{figure}
\begin{center}
\includegraphics*[width=12cm,angle=0]{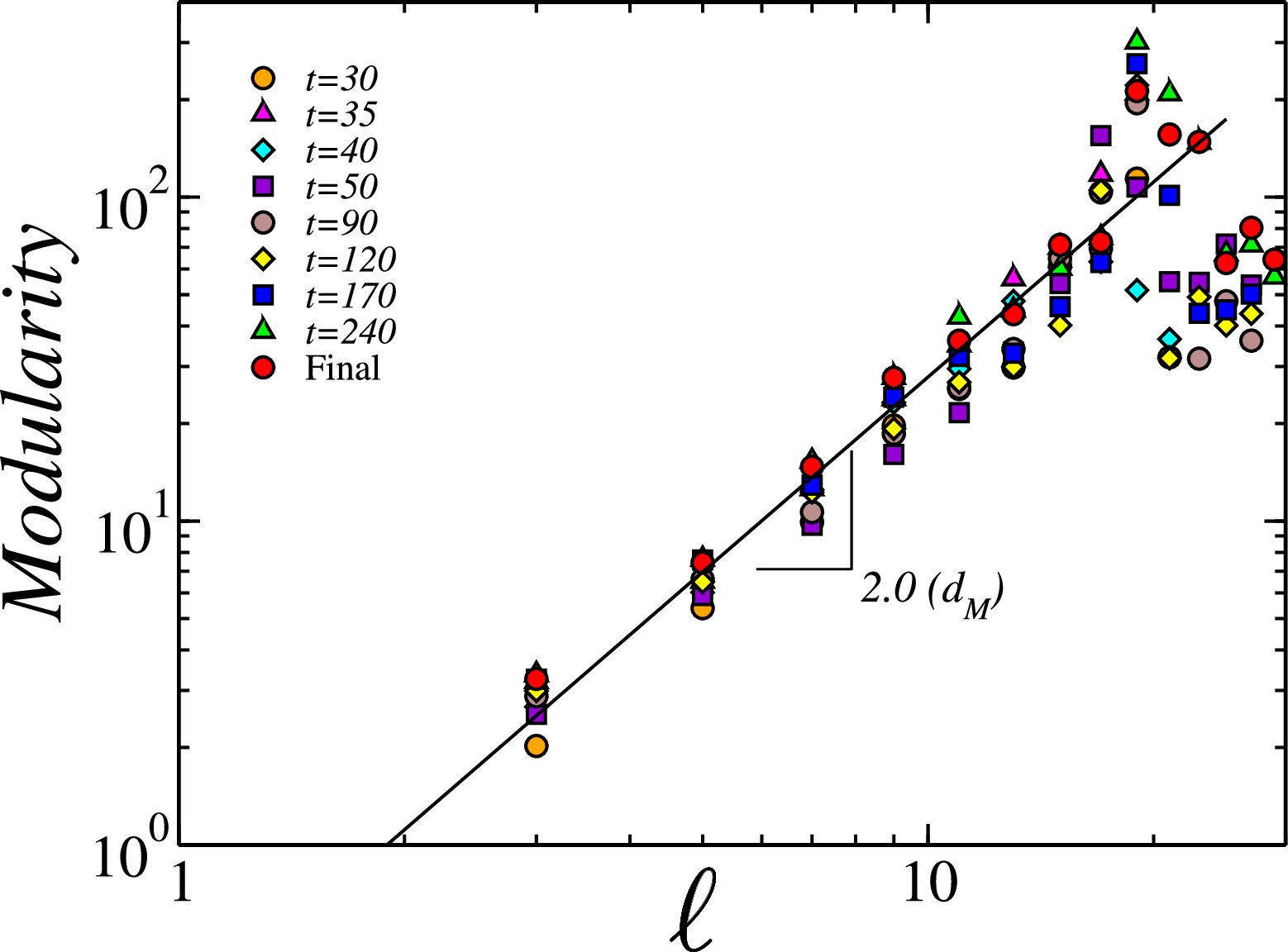}

{\bf Fig. 3a}

\vspace{1.5cm}

\includegraphics*[width=12cm,angle=0]{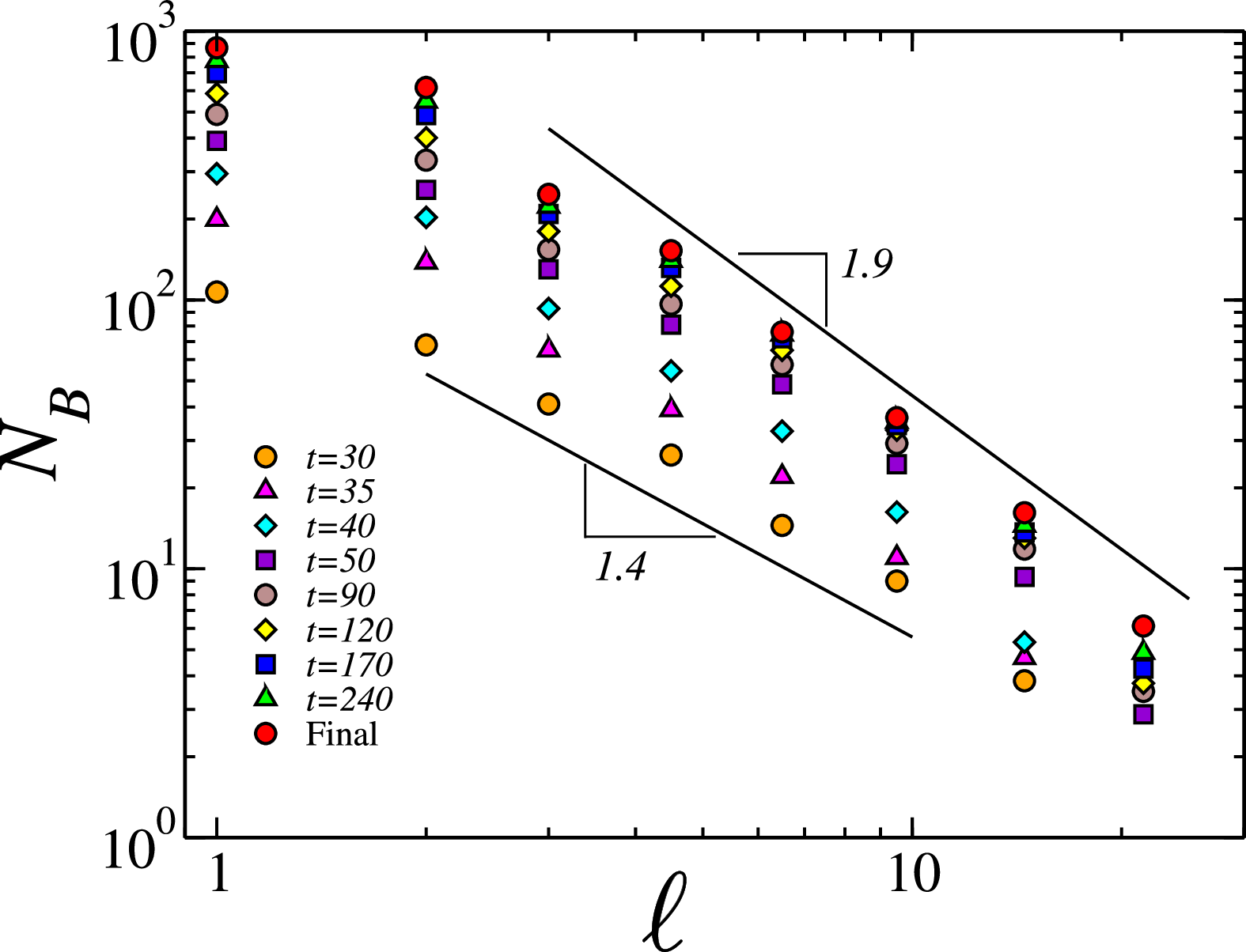}

{\bf Fig. 3b}
\end{center}
\caption{}
\label{fig7}
\end{figure}

\clearpage

\begin{figure}
\begin{center}
{\bf a} \includegraphics*[width=8cm,height=5.5cm,angle=0]{f4a.eps}
{\bf b} \includegraphics*[width=7cm,height=5.5cm,angle=0]{f8.eps}

{\bf c} \includegraphics*[width=8cm,height=5.5cm,angle=0]{f4b.eps}
{\bf d} \includegraphics*[width=7cm,height=5.5cm,angle=0]{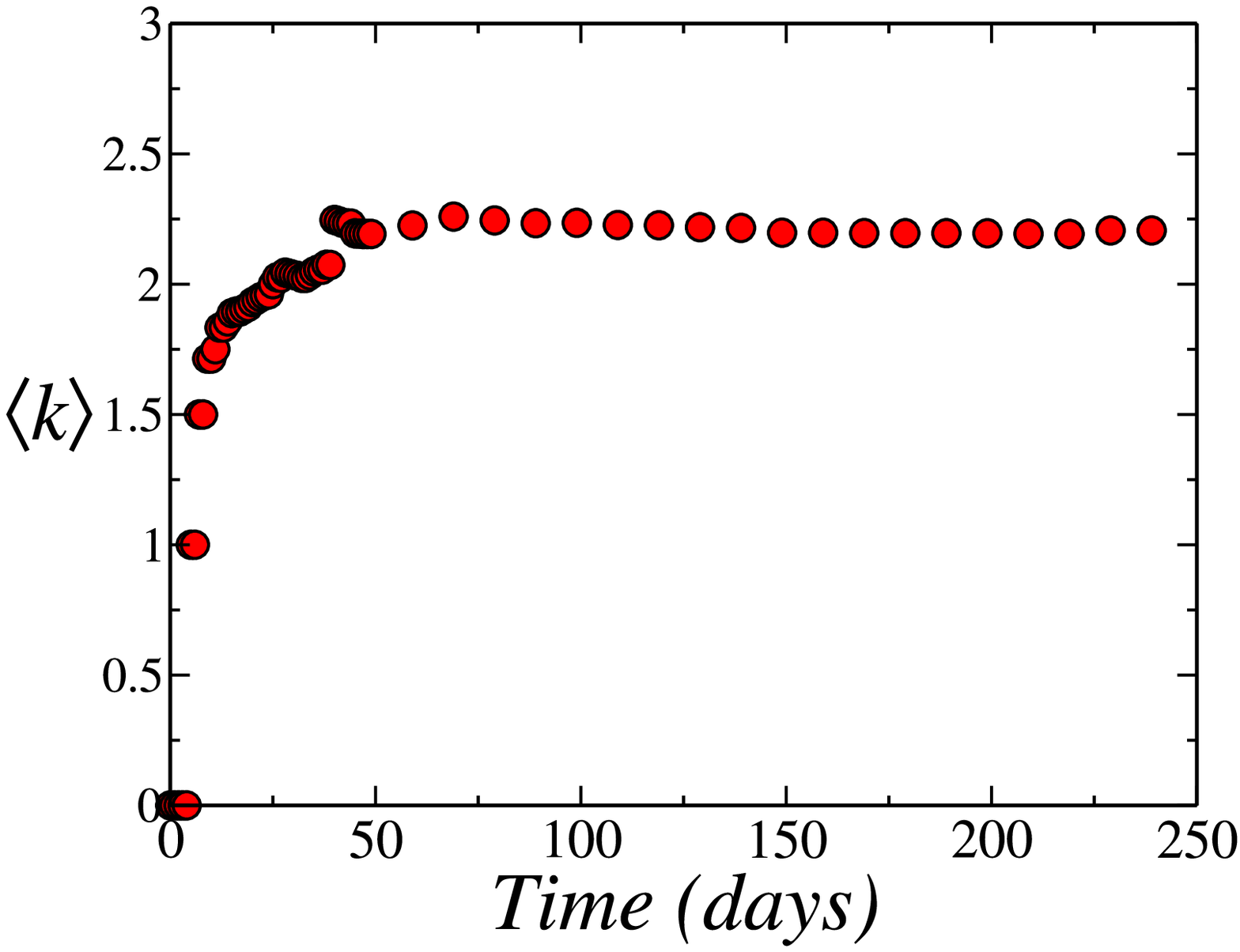}
\end{center}
\caption{}
\label{fig4}
\end{figure}

\clearpage

\centerline{\bf SUPPLEMENTARY INFORMATION}

\vspace{.5cm} \centerline{ \bf Modularity map of the network of human
  cell differentiation}

\vspace{.5cm}

\section{\bf ADDITIONAL INFORMATION FOR NHCD}
\label{additional}

{\bf FIGURE \ref{fig3}} provides an alternative representation of the
NHCD.  Different cell types appearing as bifurcation points of the
network are stacked along the vertical axis.  In SI-Fig. \ref{fig3}a
the horizontal axis corresponds to the shortest path $\ell_{N1}$
calculated from node $N1$, the fertilized egg, to any given node,
while in SI-Fig.~\ref{fig3}b the same network is shown as a function
of the node appearance time $T_{a}$.  The links emerging from each
cell type follow a bifurcation pattern ending up at the right side
with $k-1$ branches, each one of them representing one of the more
specialized cell types.  Red nodes correspond to the surviving cell
types, and they preferentially appear at later times, while the
non-surviving cell types, the blue nodes, emerge during the early
stages of the process.  The color of the edges corresponds to one of
the 19 functional groups identified in Table \ref{table1} as given in
Fig. \ref{fig1}. Links that generate loops are plotted in red.

Figure \ref{fig3}b contains the same information as Fig. \ref{fig3}a but
we plot each cell type according to its time of appearance rather than as
a function of the chemical distance to $N1$, as in Fig. \ref{fig3}a.  The
white and yellow alternating vertical stripes divide the time axis in
intervals in days. The branches have been extended so that each cell
appears only in the corresponding interval. Colors and labels are the same
as in Fig. \ref{fig3}a.

The catalog
presented in Ref.~\cite{Vickaryous} reports 407 distinct cell types in
a healthy adult human body, all of which can be identified in our
network representation. Most of them occupy the end points of the 529
branches in Figs.~\ref{fig1} and \ref{fig3}. Therefore, not all tree
leaves (branch endpoints) correspond to cell types in born humans. 

The average shortest path calculated from all cell types to $N1$ is
$\langle \ell_{N1}\rangle=10.93$, as expected from the large
concentration of links in the interval $8 \leq \ell \leq 13$
(Fig. \ref{fig3}a).  


\vspace{.5cm}

{\bf FIGURE \ref{full}} shows the full modular structure of the NHCD as
detected by the box-covering algorithm at different length scales.  A
detail of this process is represented in Fig. \ref{modules} in the
main text. A list containing the nodes belonging to each module at a
given $\ell$ is contained in the file modules.txt.

\vspace{.5cm}

{\bf FIGURE \ref{fig5}} shows the degree distribution of the NHCD for
different times.

\vspace{.5cm}

{\bf Table \ref{table1}} lists the different known functional modules
of the NHCD and the respective citations to the literature.  The
complete collected data is listed in the datafile: links.txt. This
file includes the links between the cell types, their time of
appearance in days after fecundation ($T_a$), and the known functional
module they belong to.  The references to the publications reporting
each link appear in the table.  Data on the structure of individual
communities were obtained from the specialized literature.
Some functional communities are:
germ layer and extraembryonic tissue
\cite{Alberts2002,Kirschstein2001,Sadler2004,Sell2004}, digestive
system \cite{Sadler2004,Sell2004}, pharyngeal system
\cite{Freitas1999,Sadler2004,Vickaryous}, cloacal system
\cite{Sadler2004}, neural system \cite{Sell2004,Vickaryous}, eye
\cite{Sadler2004,Vickaryous}, primitive oral cavity \cite{Sadler2004},
nose \cite{Sadler2004}, skeletal system \cite{Bianco2001,Freitas1999},
skeletal muscular system \cite{Chen2003, Sell2004,Vickaryous},
hematopoietic system \cite{Janeway2001}, urogenital system
\cite{Anglani2004,Horster1999}, cardiovascular system
\cite{Sadler2004,Vickaryous}, mesothelium \cite{Herrick2004},
respiratory \cite{Otto2002}, neural crest
\cite{Jessen2005,Nakashima2003, Sadler2004,Santagati2003},
adenohypophysis \cite{Savage2003}, ear \cite{Forge2002,Vickaryous},
and integumentary system \cite{Panteleyev2001,Sadler2004,Vickaryous}.

\clearpage


\begin{figure}
\begin{center}
\includegraphics*[width=12.cm,height=17.25cm,angle=0]{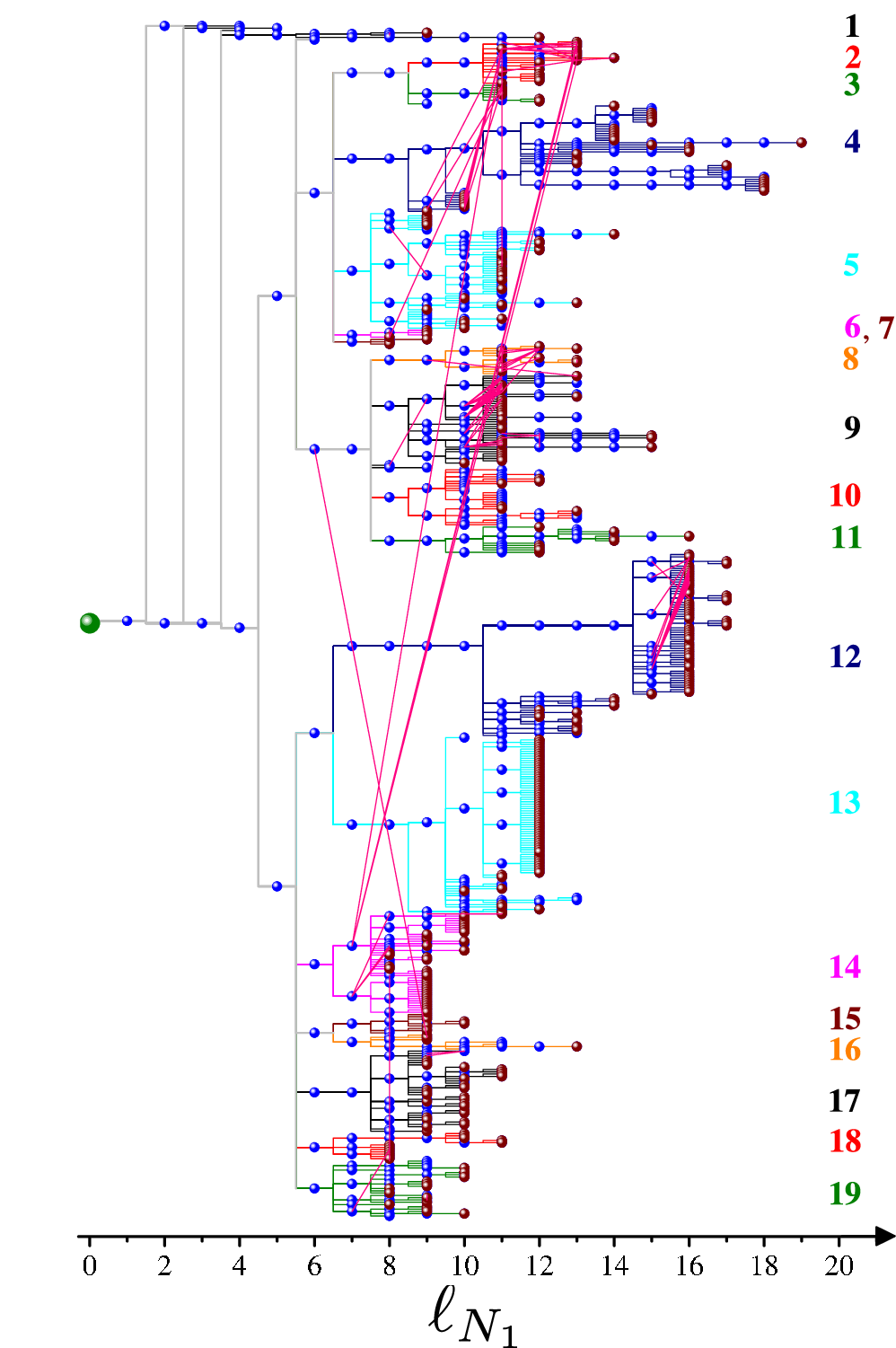}

{\bf Fig. 5a} 
\end{center}
\end{figure}
\clearpage
\begin{figure}
\begin{center}
\includegraphics*[width=12.cm,height=17.25cm,angle=0]{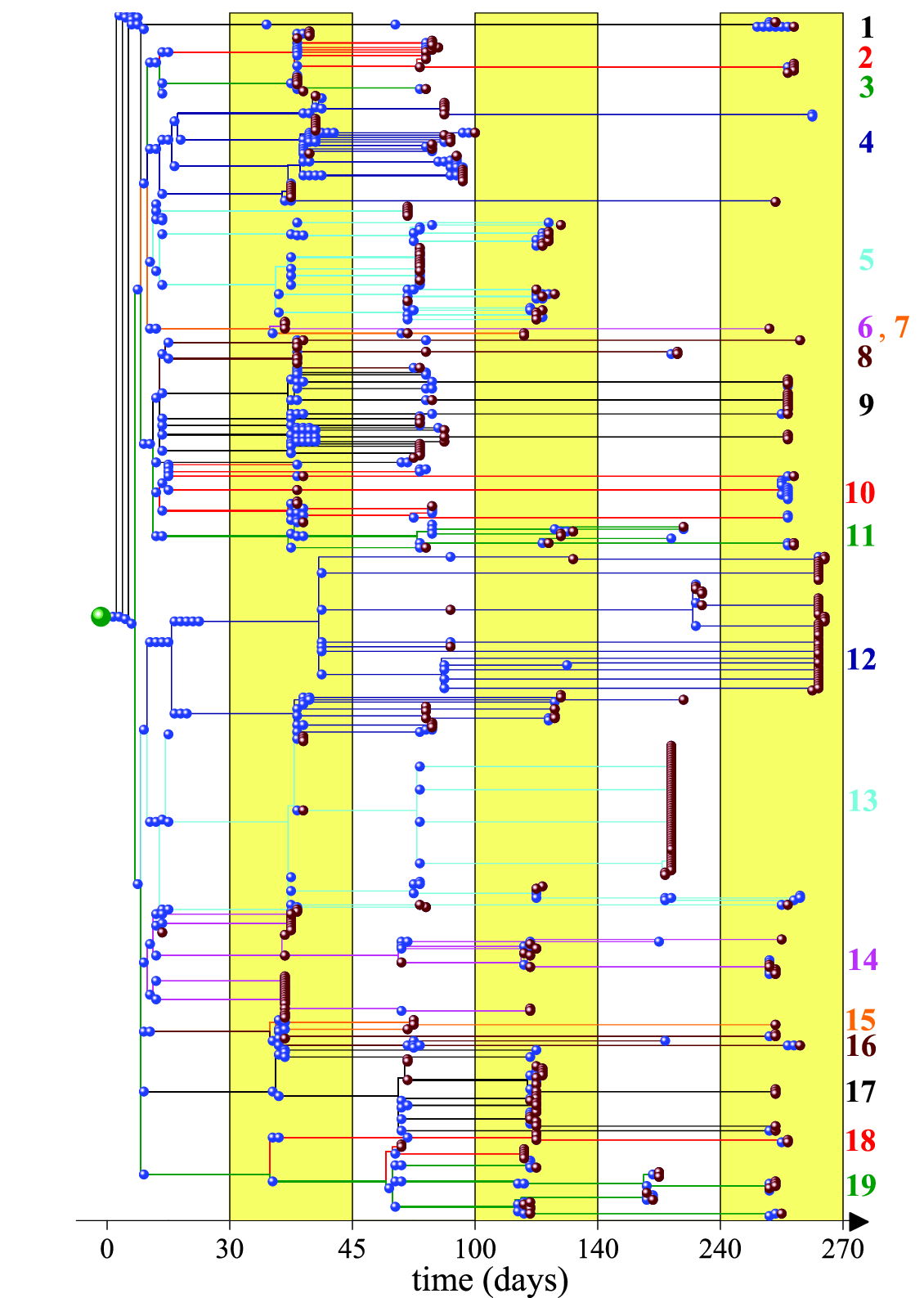}

{\bf Fig. 5b} 
\end{center}
\caption{{\bf Alternative representation of the NHCD.}  ({\bf a}) The
  horizontal axis measures the shortest path from each cell type to
  the fertilized egg along the network, $\ell_{N_1}$. ({\bf b}) The
  horizontal axis denotes the appearance time of a given cell type,
  $T_a$.  For simplicity we do not plot the links leading to loops.
  The colors of the branches denote the functional classes as in
  Fig. \ref{fig1}.  }
\label{fig3}
\end{figure}

\begin{figure}
\begin{center}
\includegraphics[height=20.cm,angle=0]{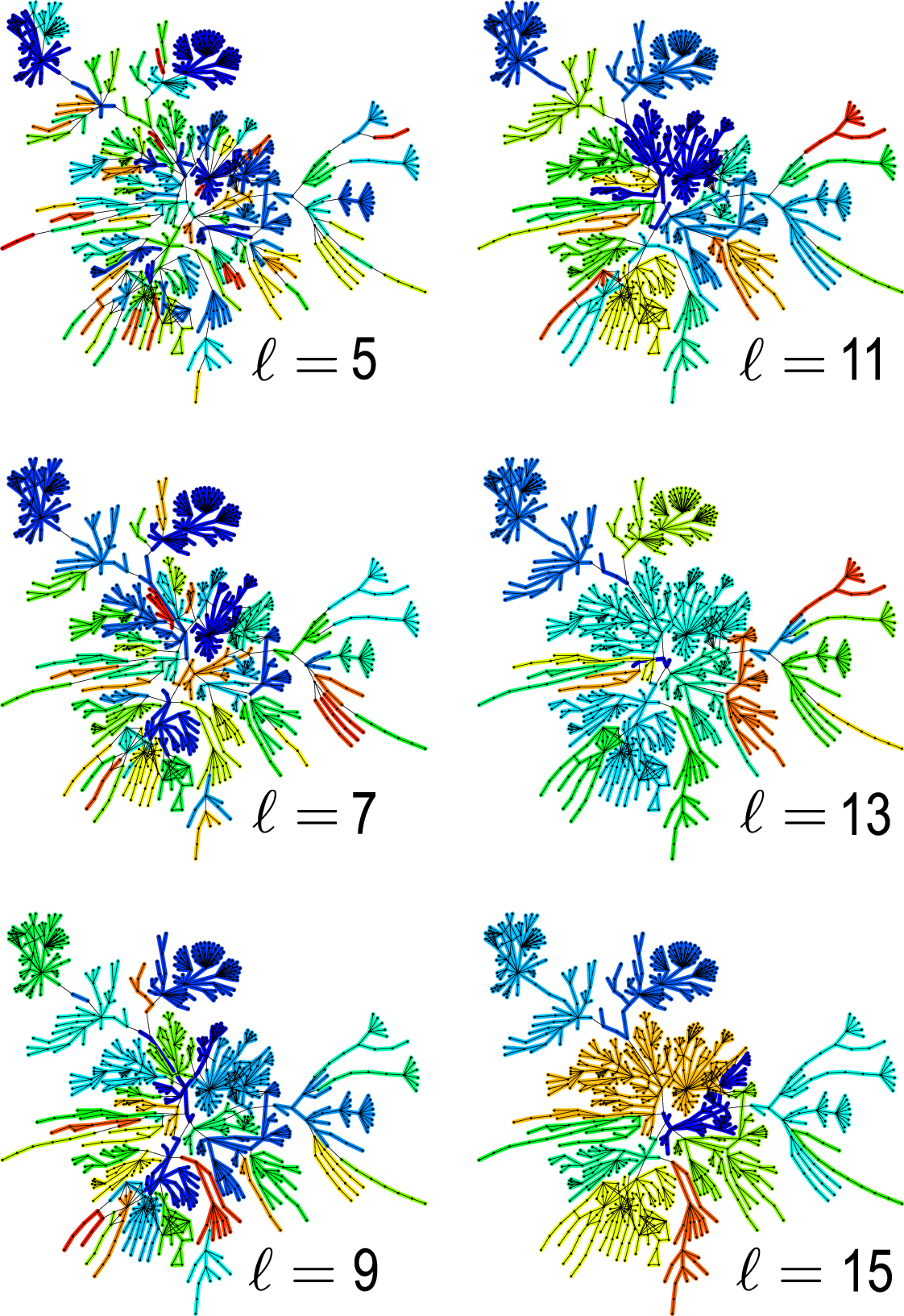}
\end{center}
\end{figure}

\begin{figure}
\begin{center}
\includegraphics[height=20.cm,angle=0]{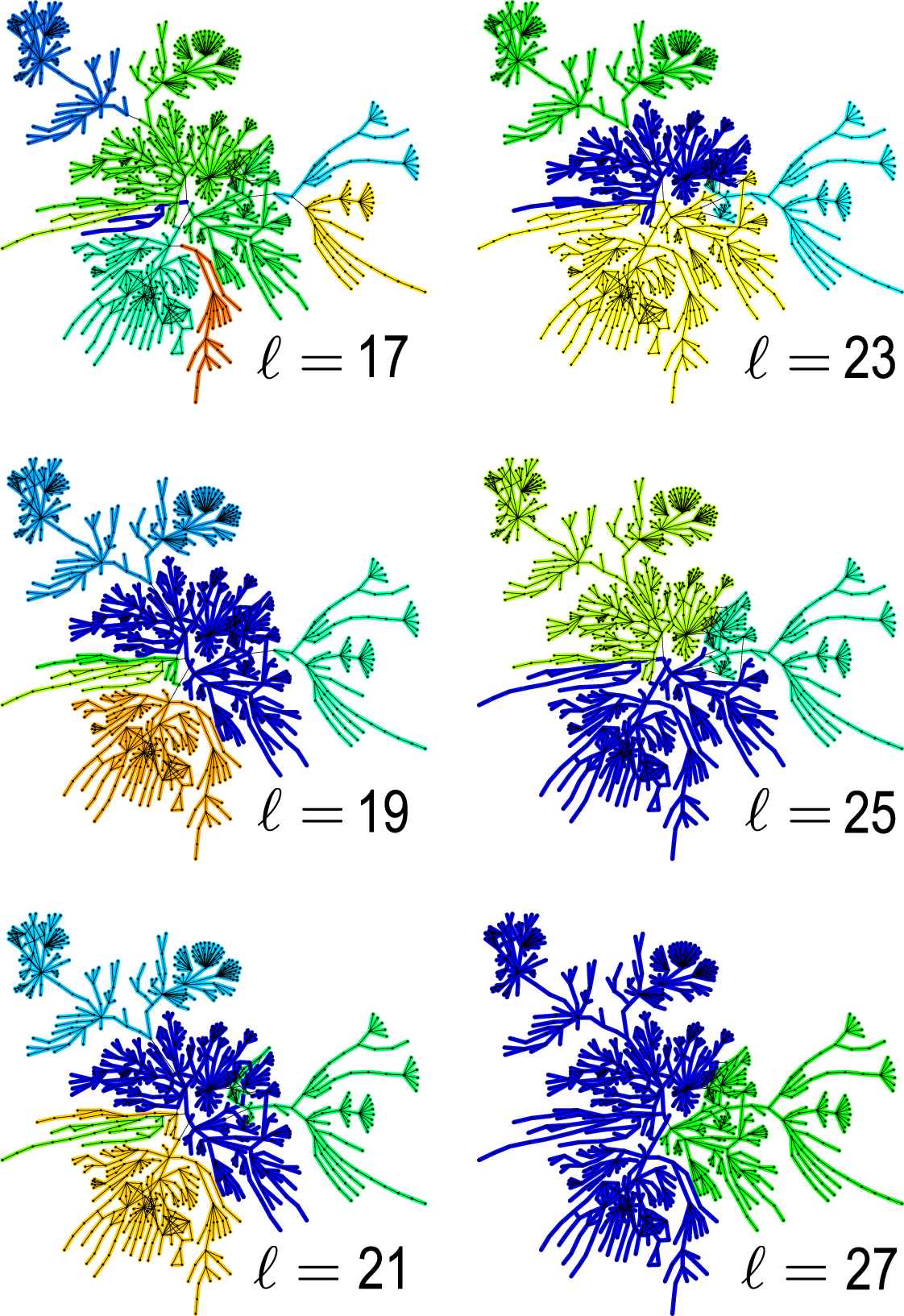}
\end{center}
\caption{{\bf Full modular structure of the NHCD at the indicated
    length $\ell$, as detected by the box-covering algorithm.}  Each node is depicted with a different color
    indicating the module to which it belongs to.
}
\label{full}
\end{figure}

\begin{figure}
\begin{center}
\includegraphics*[width=12cm,height=8.25cm,angle=0]{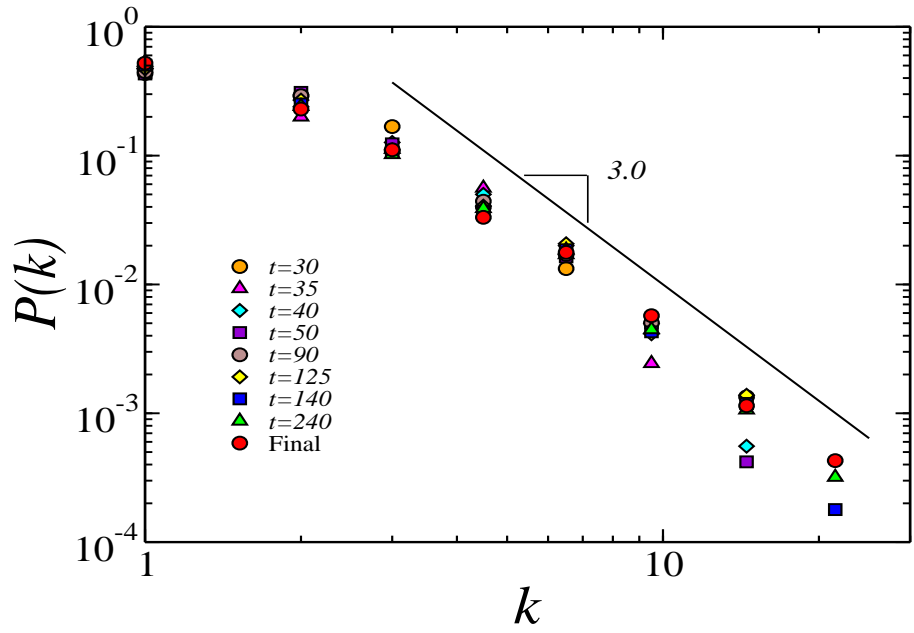}
\end{center}
\caption{Degree distribution $P(k)$ for the NHCD at different growth
  stages, from $t=30$ days to $t=240$ days.}\label{fig5}
\end{figure}

\clearpage

\begin{table}
\caption{Identification of the distinct biological functions
    of the cell types indicated in Fig. \ref{fig1} and SI-Fig. \ref{fig3}. The third column lists
    the references used for building the NHCD.} \label{table1}
\begin{tabular}{|l|l|l|}
  \hline
  Label & Biological function & Reference \\ \hline 
  1 & Germinative Lineage & Alberts et al., 2002; Kirschstein and Skirboll, 2001;\\  
  &  and Placenta &  Sadler, 2004; Sell, 2004 \\  \hline
  2 & Skeletal System & Bianco et al., 2001; Freitas, 1999; Mochida, 2005; Sadler, 2004; \\
  &  &  Sell, 2004; Sanders et al., 1999; Towler and  Gelberman, 2006;\\ 
  &  &    Vickaryous and Hall, 2006 \\  \hline
  3 & Skeletal Muscular System & Chen and Goldhamer, 2002; Sadler, 2004; Sell, 2004; \\
  &  &    Vickaryous and Hall, 2006 \\  \hline
  4 & Hematopoietic Lineage & Alberts et al., 2002; Kirschstein and Skirboll, 2001; \\
  &  &  Janeway et al., 2001; Minasi et al., 2002; Paxinos and Mai, 2004;  \\
  &  &  Sadler, 2004; Sell, 2004; Vickaryous and Hall, 2006  \\ \hline
  5 & Urogenital System  & Anglani et al., 2004; Coulter, 2004; Horster et al., 1999; Lopez\\
  &  & et al., 2001; Sadler, 2004; Sell, 2004; Vickaryous and Hall, 2006 \\ \hline
  6 & Cardiovascular System & Sadler, 2004; Sell, 2004; Vickaryous and Hall, 2006  \\ \hline
  7 & Mesothelial Lineage & Herrick and Mutsaers, 2004; Sadler, 2004  \\ \hline
  8 & Respiratory System & Freitas, 1999; Otto, 2002; Sadler, 2004; Sell, 2004  \\ \hline
  9 & Digestive System & Bardeesy and DePinho, 2002; Fausto, 2004; Freitas, 1999; \\ 
  &  & Sadler, 2004; Sell, 2004; Vickaryous and Hall, 2006 \\ \hline
  10 & Pharyngeal Lineage & Blackburn  and Manley, 2004; Freitas, 1999; \\
  &  & Sadler, 2004; Vickaryous and Hall, 2006  \\ \hline
  11 & Cloacal Lineage & Foster et al., 2002; Freitas, 1999; Sadler, 2004; \\
  &  & Sell, 2004; Vickaryous and Hall, 2006  \\ \hline
  12 & Neural Lineage & Freitas, 1999; Kirschstein and Skirboll, 2001; Paxinos and Mai, 2004;\\
  &  & Sadler, 2004; Sell, 2004; Temple, 2001; Vickaryous and Hall, 2006  \\ \hline
  13 & Eye Lineage & Paxinos and Mai, 2004; Sadler, 2004; Sell, 2004; \\
  &  &  Vickaryous and Hall, 2006 \\ \hline
  14 & Neural Crest Lineage & Jessen and Mirsky, 2005; Nakashima and Redid, 2003; \\
  &  &   Sadler, 2004; Santagati and Rijli, 2003; Sell, Szeder et al., 2003; \\
  &  &   Vickaryous and Hall, 2006  \\ \hline
  15 & Adenohypophysis & Paxinos and Mai, 2004; Sadler, 2004; Savage et al., 2003; \\
  &  &  Vickaryous and Hall, 2006 \\
  \hline
\end{tabular}
\end{table}

\begin{table}
\begin{tabular}{|l|l|l|}
  \hline
  16 & Primitive Oral Cavity & Freitas, 1999; Nakashima and Redid, 2003;  Sadler, 2004;\\
  &  &  Vickaryous and Hall, 2006  \\
  \hline
  17 & Ear & Forge and Wright, 2002; Freitas, 1999; Paxinos and Mai, 2004; \\
  &  &   Sadler, 2004; Vickaryous and Hall, 2006  \\  \hline
  18 & Nose & Freitas, 1999; Sadler, 2004; Vickaryous and Hall, 2006 \\  \hline
  19 & Integumentary System & Freitas, 1999; Hennighausen and Robinson, 2005; \\
  &  &   Panteleyev et al., 2001; Potten and Booth, 2002; Sadler, 2004; \\
  &  &   Stoeckelhuber et al., 2003; Vickaryous and Hall, 2006  \\
    \hline
  \end{tabular}
\end{table}

\clearpage

1.  Alberts, B., Johnson, A., Lewis, J., Raff, M., Roberts, K., \& Walter,
P. {\it Molecular Biology of the Cell} (Fourth ed., Garland Science, New York,
2002).

2.  Anglani, F., Forino, M., Del Prete, D., Tosetto, E., Torregrossa, R.,
\& D'Angelo, A. In search of adult renal stem cells. \textit{J.
Cell. Mol. Med.} \textbf{8}, 474-487 (2004).

3.  Bardeesy, N., \& DePinho, R.A. Pancreatic cancer biology and
genetics. {\it Nat. Rev. Cancer} {\bf 2}, 897-909 (2002).

4.  Bianco, P., Riminucci, M., Gronthos, S., \& Robey, P.G. Bone
marrow stromal stem cells: nature, biology, and potential applications.
{\it Stem Cells} {\bf 19}, 180-192 (2001).

5.  Blackburn, C.C., \& Manley, N.R. Developing a new paradigm for
thymus organogenesis.  {\it Nat. Rev. Immunol.} {\bf 4}, 278-289 (2004).

6.  Chen, J.C.J., \& Goldhamer, D.J. Skeletal muscle stem cells.
{\it Reprod. Biol. Endocrinol.} {\bf 1}, 101 (2003).

7.  Coulter, C.L. Functional biology of the primate fetal adrenal
gland: advances in technology provide new insight. {\it Clin. Exp. Pharmacol.}
{\bf 31}, 475-484 (2004).

8.  Fausto, N. Liver regeneration and repair: hepatocytes,
progenitor cells, and stem cells. {\it Hepatology} {\bf 39}, 1477-1487 (2004).

9.  Freitas, R.A., Jr. {\it Nanomedicine, Volume I: Basic Capabilities}
(Landes Bioscience, Georgetown, Texas, 1999).

10. Forge, A. \&  Wright, T. The molecular architecture of the inner
ear. {\it Brit. Med. Bull.} {\bf 63}, 5-24 (2002).

11. Foster, C.S., Dodson, A., Karavana, V., Smith, P.H., \& Ke, Y.
Prostatic stem cells. {\it J. Pathol.} {\bf 197}, 551-565 (2002).

12. Hennighausen, L. \& Robinson, G.W. Information networks in the
mammary gland. {\it Nat. Rev. Mol. Cell Bio.} {\bf 6}, 715-725 (2005).

13. Herrick, S.E. \& Mutsaers, S.E. Mesothelial progenitor cells and
their potential in tissue engineering. {\it Int. J. Biochem. Cell B.} {\it 36},
621-642 (2004).

14. Horster, M.F., Braun, G.S., \& Huber, S.M. Embryonic renal
epithelial: induction, nephrogenesis, and cell differentiation. {\it Physiol.
Rev.} {\bf 79}, 1157-1191 (1999).

15. Kirschstein, R. \& Skirboll, L. R. {\it Stem Cells: Scientific
Progress and Future Research Directions} (NIH, Bethesda, 2001).

16. Janeway, C.A., Travers, P., Walport, M., \& Shlomchik, M.
{\it Immunobiology: the immune system in health and disease} (fifth ed., Garland
Science, 2001).

17. Jessen, K.R. \& Mirsky, R. The origin and development of glial
cells in peripheral nerves.  {\it Nat. Rev. Neurosci.} {\bf 6}, 671- 682 (2005).

18. Lopez, M.L.S.S, Pentz, E.S., Robert, B., Abrahamson, D.R., \& Gomez, R.A.
Embryonic origin and lineage of juxtaglomerular cells. {\it Am. J.
Physiol. Renal Physiol.} {\bf 281}, 345-356 (2001).

19. Minasi MG, Riminucci M, De Angelis L, Borello U, Berarducci B,
Innocenzi A, Caprioli A, Sirabella D, Baiocchi M, De Maria R, Boratto R,
Jaffredo T, Broccoli V, Bianco P, \& Cossu G. The meso-angioblast: a
multipotent, self-renewing cell that originates from the dorsal aorta and
differentiates into most mesodermal tissues. {\it Development} {\bf 129}, 2773-2783 (2002).

20. Mochida, J. New strategies for disc repair: novel preclinical
trials. {\it J. Orthop. Sci.} {\bf 10}, 112-118 (2005).

21. Nakashima, M. \& Redid, A.H. The application of bone
morphogenetic proteins to dental tissue engineering. {\it Nat. Biotechnol.} {\bf 21},
1025-1032 (2003).

22. Otto, W.R. Lung epithelial stem cells. {\it J. Pathol.} {\bf 197},
527-535 (2002).

23. Panteleyev, A., Jahoda, C.A.B., \& Christiano, A.M., Hair follicle
predetermination. {\it J. Cell. Sci.} {\bf 114}, 3419-3431 (2001).

24. Paxinos, G. \& Mai J.K. {\it The Human Nervous System} (second ed.,
Elsevier Academic Press, 2004).

25. Potten, C.S. \& Booth, C. Keratinocyte Stem Cells: a Commentary.
{\it J. Invest. Detmatol.} {\it 119}, 888-899 (2002).

26. Sadler, T.W. {\it Langman's Medical Embryology} (ninth ed.,
Lippincott Williams \& Wilkins, Baltimore, 2004).

27. Santagati, F. \& Rijli, F.M. Cranial neural crest and the
building of the vertebrate head. {\it Nat. Rev. Neurosci.} {\bf 4}, 806-818 (2003).

28. Savage, J.J., Yaden, B.C. Kiratipranon, P., \& Rhodes, S.J.
Transcriptional control during mammalian anterior pituitary development.
{\it Gene} {\bf 319}, 1-19 (2003).

29. Sell, S. {\it Stem Cells Handbook} (Humana Press, Totowa, NJ, 2004).

30. Stoeckelhuber, M., Stoeckelhuber, B.M., \& Welsch, U. Human Glands
of Moll: Histochemical and Ultrastructural Characterization of the Glands
of Moll in the Human Eyelid. {\it J. Invest. Dermatol.} {\bf 121}, 28-36 (2003).

31. Szeder, V., Grim, M., Halata, Z., \& Sieber-Bluma, M. Neural crest
origin of mammalian Merkel cells. {\it Dev. Biol.} {\bf 253}, 258-263 (2003).

32. Temple, S. The development of neural stem cells. {\it Nature}, {\bf 414},
112-117 (2001).

33. Towler, D.A. \& Gelberman, R.H. The alchemy of tendon repair: a
primer for the (S)mad scientist. {\it J. Clin. Invest.} {\bf 116}, 863-866 (2006).

34. Vickaryous, M.K. \& Hall, B.K. Human cell type diversity,
evolution, development, and classification with special reference to cells
derived from the neural crest. {\it Biol. Rev.} {\bf 81}, 425-455 (2006).

\end{document}